\def\be{\begin{equation}}
\def\ee{\end{equation}}
\def\ba{\begin{array}}
\def\ea{\end{array}}

\def\Nb{{I\!\! N}}

\def\Cb{\ \hbox{\vrule width 0.6pt height 6pt depth 0pt
              \hskip -3.2 pt} C}
\documentstyle[12pt]{article}
\begin{document}
\parskip=4pt
\parindent=18pt
\baselineskip=22pt \setcounter{page}{1} \centerline{\Large\bf A
Class of Special Matrices and Quantum Entanglement} \vspace{6ex}
\begin{center}
Shao-Ming Fei$^{\dag\ddag}$~~~ and ~~~Xianqing Li-Jost$^\ddag$
\bigskip

\begin{minipage}{6.2in}
$^\dag$Department of Mathematics, Capital Normal University, Beijing 100037, P.R. China\\
$^\dag$Institut f{\"u}r Angewandte Mathematik, Universit{\"a}t Bonn, 53115 Bonn, Germany\\
E-mail: fei@wiener.iam.uni-bonn.de\\
$^\ddag$Max-Planck-Institute for Mathematics in the Sciences, 04103 Leipzig, Germany\\
E-mail: Xianqing.Li-Jost@mis.mpg.de

\end{minipage}
\end{center}
\vskip 1 true cm
\parindent=18pt
\parskip=6pt
\begin{center}
\begin{minipage}{5in}
\vspace{3ex} \centerline{\large Abstract} \vspace{4ex}

We present a kind of construction for a class of special matrices
with at most two different eigenvalues, in terms of some
interesting multiplicators which are very useful in calculating
eigenvalue polynomials of these matrices. This class of matrices
defines a special kind of quantum states --- $d$-computable
states. The entanglement of formation
for a large class of quantum mixed states is explicitly presented.

\end{minipage}
\end{center}


\smallskip
\bigskip

\section{Introduction}

Quantum entangled states are playing an important role in quantum
communication, information processing and quantum computing
\cite{DiVincenzo}, especially in the investigation of quantum
teleportation \cite{teleport,teleport1}, dense coding
\cite{dense}, decoherence in quantum computers and the evaluation
of quantum cryptographic schemes \cite{crypto}. To quantify
entanglement, a number of entanglement measures such as the
entanglement of formation and distillation
\cite{Bennett96a,BBPS,Vedral}, negativity
\cite{Peres96a,Zyczkowski98a}, relative entropy \cite{Vedral,sw}
have been proposed for bipartite states \cite{crypto,BBPS}
[11-13]. Most of these measures of entanglement involve
extremizations which are difficult to handle analytically. For
instance the entanglement of formation \cite{Bennett96a} is
intended to quantify the amount of quantum communication required
to create a given state. The entanglement of formation for
a pair of qubits can be
expressed as a monotonically increasing function of the
``concurrence'', which can be taken as a measure of entanglement
in its own right \cite{HillWootters}. From the expression of this
concurrence, the entanglement of formation for mixed states of a
pair of qubits is calculated \cite{HillWootters}. Although
entanglement of formation is defined for arbitrary dimension, so
far no explicit analytic formulae for entanglement of formation
have been found for systems larger than a pair of qubits, except
for some special symmetric states \cite{th}.

For a multipartite quantum system, the degree of entanglement will
neither increase nor decrease under local unitary transformations
on a quantum subsystem. Therefore the measure of entanglement must
be an invariant of local unitary transformations. The
entanglements have been studied in the view of this kind of
invariants and a generalized formula of concurrence for high
dimensional bipartite and multipartite systems is derived from the
relations among these invariants \cite{note}. The generalized
concurrence can be used to deduce necessary and sufficient
separability conditions for some high dimensional mixed states
\cite{qsep}. However in general the generalized concurrence is not
a suitable measure for $N$-dimensional bipartite quantum pure
states, except for $N=2$. Therefore it does not help in calculating the
entanglement of formation for bipartite mixed states.

Nevertheless in \cite{fjlw} it has been shown that for
some class of quantum states with $N>2$,
the corresponding entanglement of formation is
a monotonically increasing function of a generalized
concurrence, and the entanglement of formation can be also calculated
analytically. Let ${\cal H}$ be an $N$-dimensional complex Hilbert space with
orthonormal basis $e_i$, $i=1,...,N$. A general bipartite pure
state on ${\cal H}\otimes {\cal H}$ is of the form,
\begin{equation}\label{psi}
\vert\psi>=\sum_{i,j=1}^N a_{ij}e_i\otimes e_j,~~~~~~a_{ij}\in\Cb
\end{equation}
with normalization $\displaystyle\sum_{i,j=1}^N a_{ij}a_{ij}^\ast=1$.
The entanglement of formation $E$ is defined as the entropy of
either of the two sub-Hilbert spaces \cite{BBPS},
\be
\label{epsiq}
E(|\psi \rangle) = - {\mbox{Tr}\,}
(\rho_1 \log_2 \rho_1) = - {\mbox{Tr}\,} (\rho_2 \log_2 \rho_2)\,,
\ee
where $\rho_1$ (resp. $\rho_2$) is the partial trace of $\bf
|\psi\rangle\langle\psi|$ over the first (resp. second) Hilbert
space of ${\cal H}\otimes{\cal H}$.
Let $A$ denote the matrix with
entries given by $a_{ij}$ in (\ref{psi}). $\rho_1$ can be
expressed as $\rho_1=AA^\dag$.

The quantum mixed states are described by
density matrices $\rho$ on ${\cal H}\otimes{\cal H}$,
with pure-state decompositions, i.e., all ensembles of states $|\psi_i \rangle$
of the form (\ref{psi}) with probabilities $p_i\geq 0$,
$\rho = \sum_{i=1}^l p_i |\psi_i \rangle \langle\psi_i|$,
$\sum_{i=1}^l p_i =1$
for some $l\in\Nb$. The entanglement of formation for the mixed
state $\rho$ is defined as the average entanglement of the pure
states of the decomposition, minimized over all decompositions of
$\rho$, $E(\rho) = \mbox{min}\, \sum_{i=1}^l p_i
E(|\psi_i \rangle)$.

For $N=2$ equation (\ref{epsiq}) can be written as
$
E(|\psi \rangle)|_{N=2} =h((1+\sqrt{1-C^2})/2),
$
where $h(x) = -x\log_2 x - (1-x)\log_2 (1-x)$.
$C$ is called concurrence,
$C(|\psi \rangle) =2|a_{11}a_{22}-a_{12}a_{21}\vert$ \cite{HillWootters}.
$E$ is a monotonically increasing function of $C$ and therefore $C$ can be
also taken as a kind of measure of entanglement. Calculating
$E(\rho)$ is then reduced to the calculation of the corresponding minimum of
$C(\rho) = \mbox{min}\, \sum_{i=1}^M p_i C(|\psi_i \rangle)$,
which simplifies the problems, as $C(|\psi_i \rangle)$ has a much simpler
expression than $E(|\psi_i \rangle)$.

For $N\geq 3$, there is no such concurrence $C$ in general. The
concurrences discussed in \cite{note} can be only used to judge
whether a pure state is separable (or maximally entangled) or not
\cite{qsep}. The entanglement of formation is no longer a
monotonically increasing function of these concurrences.

Nevertheless, for a special class of quantum states
such that $AA^\dag$ has only two non-zero eigenvalues, a kind of
generalized concurrence has been found to simplify the
calculation of the corresponding entanglement of formation \cite{fjlw}.
Let $\lambda_1$ (resp. $\lambda_2$) be the two non-zero
eigenvalues of $AA^\dag$ with degeneracy $n$ (resp. $m$), $n+m\leq N$,
and $D$ the maximal non-zero diagonal determinant, $D=\lambda_1^n\lambda_2^m$.
In this case the entanglement of formation of $|\psi \rangle$ is given by
$E(|\psi \rangle)=-n \lambda_1 \log_2 \lambda_1 -m \lambda_2 \log_2 \lambda_2$.
It is straightforward to show that $E(|\psi\rangle)$ is a
monotonically increasing function of $D$ and hence
$D$ is a kind of measure of entanglement in this case.
In particular for the case $n=m>1$, we have
\be\label{hehe}
E(|\psi \rangle)=n \left(-x\log_2 x
- (\frac{1}{n}-x)\log_2 (\frac{1}{n}-x)\right),
\ee
where
$$
x = \frac{1}{2}\left(\frac{1}{n}+\sqrt{\frac{1}{n^2}(1-d^2)}\right)
$$
and
\be\label{GC}
d\equiv 2nD^{\frac{1}{2n}}=2n\sqrt{\lambda_1\lambda_2}.
\ee
$d$ is defined to be the generalized concurrence in this case.
Instead of calculating $E(\rho)$ directly, one may calculate the
minimum decomposition of $D(\rho)$ or $d(\rho)$ to simplify the
calculations.
In \cite{fjlw} a class of pure states (\ref{psi}) with the matrix $A$ given by
\be\label{a} A=\left( \ba{cccc}
0&b&a_1&b_1\\
-b&0&c_1&d_1\\
a_1&c_1&0&-e\\
b_1&d_1&e&0
\ea
\right),
\ee
$a_1,b_1,c_1,d_1,b,e\in\Cb$, is considered.
The matrix $AA^\dag$ has two eigenvalues with degeneracy two, i.e., $n=m=2$ and
$\vert AA^\dag\vert=|b_1c_1-a_1d_1+be|^4$.
The generalized concurrence $d$ is given by
$d=4|b_1c_1-a_1d_1+be|$.
Let $p$ be a $16\times 16$ matrix with only non-zero entries
$p_{1,16}=p_{2,15}=-p_{3,14}=p_{4,10}=p_{5,12}=p_{6,11}
=p_{7,13}=-p_{8,8}=-p_{9,9}=p_{10,4}=p_{11,6}=p_{12,5}
=p_{13,7}=-p_{14,3}=p_{15,2}=p_{16,1}=1$.
$d$ can be further written as
\be\label{dp}
d=|\langle \psi |p\psi^* \rangle |.
\ee
Let $\Psi$ denote the set of pure states (\ref{psi})
with $A$ given as the form of (\ref{a}). Consider all mixed states with
density matrix $\rho$ such that its decompositions are of the form
\be\label{rho1}
\rho = \sum_{i=1}^M p_i |\psi_i \rangle \langle \psi_i|,~~~~\sum_{i=1}^M
p_i =1,~~~~|\psi_i\rangle\in\Psi.
\ee
All other kind of decompositions, say decomposition from $|\psi_i^\prime\rangle$,
can be obtained from a unitary linear combination
of $|\psi_i\rangle$ \cite{HillWootters,fjlw}. As linear combinations of
$|\psi_i\rangle$ do not change the form of the corresponding matrices (\ref{a}),
once $\rho$ has a decomposition with all $|\psi_i\rangle\in\Psi$, all
other decompositions, including the minimum decomposition of the entanglement
of formation, also satisfy that $|\psi_i^\prime\rangle\in\Psi$.
Then the minimum decomposition of the generalized concurrence is
\cite{fjlw}
\be \label{drho}
d(\rho)=\Lambda_1 - \sum_{i=2}^{16}\Lambda_i,
\ee
where $\Lambda_i$, in decreasing
order, are the square roots of the eigenvalues of the Hermitian
matrix $R \equiv \sqrt{\sqrt{\rho}p{\rho^\ast}p\sqrt{\rho}}$, or,
alternatively, the square roots of the eigenvalues of the
non-Hermitian matrix $\rho p{\rho^\ast}p$.

\section{Entanglement of formation for a class of high dimensional
quantum states}

An important fact in obtaining the formula (\ref{drho}) is that
the generalized concurrence $d$ is a quadratic form of the entries of
the matrix $A$, so that $d$ can be expressed in the form of (\ref{dp})
in terms of a suitable matrix $p$. Generalizing to the $N$-dimensional
case we call an pure state (\ref{psi})
\underline{$d$-computable} if $A$ satisfies the following relations:
\be\label{dcomputable}
\ba{l}
\vert AA^\dag\vert = ([ A ][ A ]^\ast)^{N/2},\\[3mm]
\vert AA^\dag - \lambda Id_N\vert=(\lambda^2 - \| A \| \lambda + [ A ][ A ]^\ast)^{N/2},
\ea
\ee
where $[A]$ and $\|A \|$ are quadratic forms of $a_{ij}$,
$Id_N$ is the $N\times N$ identity matrix.
We denote ${\cal A}$ the set of matrices satisfying (\ref{dcomputable}),
which implies that for $A\in{\cal A}$, $AA^\dag$ has at most
two different eigenvalues, each one has order $N/2$ and
$d$ is a quadratic form of the entries of the matrix $A$.

In the following we give a kind of constructions of high dimensional
$d$-computable states. For all $N^2\times N^2$
density matrices with decompositions on these $N$-dimensional
$d$-computable pure states, their entanglement of formations
can be calculated with a similar formula to (\ref{drho})
(see (\ref{d2k1})).

We first present a kind of construction for a class of
$N$-dimensional, $N = 2^k$, $2\leq k\in\Nb$,
$d$-computable states.
Set
$$
A_2= \left(
\begin{array}{cc}
a&-c \\[3mm]
c&d\\[3mm]
\end{array}
\right),
$$
where $a,c,d \in \Cb$.
For any $b_1,c_1 \in \Cb$, a $4\times 4$ matrix $A_4\in{\cal A}$ can
be constructed in the following way,
\be\label{ha4}
A_4= \left(
\begin{array}{cc}
B_2&A_2\\[3mm]
-A_2^t&C_2^t\\[3mm]
\end{array}
\right) = \left(
\begin{array}{cccc}
0&b_1&a&-c\\[3mm]
-b_1&0&c&d\\[3mm]
-a&-c&0&-c_1\\[3mm]
c&-d&c_1&0
\end{array}
\right),
\ee
where
$$
B_2 = b_1J_2, ~~~~ C_2 = c_1J_2, ~~~ J_2= \left(
\begin{array}{cc}
0&1 \\[3mm]
-1&0\\[3mm]
\end{array}
\right).
$$
$A_4$ satisfies the relations in (\ref{dcomputable}):
$$
\begin{array}{l}
\left| A_4 A^\dag_4 \right|=[(b_1c_1+a d + c^2)(b_1c_1+a d + c^2)^\ast]^2=
([ A_4 ][ A_4 ]^\ast)^2,\\[3mm]
\left| A_4 A^\dag_4 - \lambda Id_4 \right| = (\lambda^2 -
(b_1b_1^\ast+c_1c_1^\ast+aa^\ast+2cc^\ast+dd^\ast)\lambda\\[3mm]
~~~~~~~~~~~~~~~~~~~~~~+ (b_1c_1+ a d + c^2)(b_1c_1+ a d + c^2)^\ast)^2\\[3mm]
~~~~~~~~~~~~~~~~~~~= (\lambda^2 - \| A_4 \|\lambda + [ A_4 ][ A_4 ]^\ast)^2,
\end{array}
$$
where
\be
[ A_4 ]=(b_1c_1+a d + c^2),~~~
\| A_4 \|=b_1b_1^\ast+c_1c_1^\ast+aa^\ast+2cc^\ast+dd^\ast.
\ee

$A_8\in{\cal A}$ can be obtained from $A_4$,
\be\label{a8} A_8=
\left(
\begin{array}{cc}
B_4&A_4\\[3mm]
-A_4^t&C_4^t\\[3mm]
\end{array}
\right), \ee where
\be\label{i4} B_4 = b_2J_4, ~~~~C_4 = c_2J_4,
~~~~ J_4= \left(
\begin{array}{cccc}
0&0&0&1\\[3mm]
0&0&1&0\\[3mm]
0&-1&0&0\\[3mm]
-1&0&0&0
\end{array}
\right),~~~~
b_2,~c_2 \in \Cb.
\ee

For general construction of high dimensional matrices
$A_{2^{k+1}}\in{\cal A}$, $2 \leq k\in\Nb$, we have
\be\label{a2k}
A_{2^{k+1}}= \left(
\begin{array}{cc}
B_{2^{k}}&A_{2^{k}}\\[3mm]
(-1)^{\frac{k(k+1)}{2}}A_{2^{k}}^t&C_{2^{k}}^t
\end{array}
\right) \equiv\left(
\begin{array}{cc}
b_{k}J_{2^{k}}&A_{2^{k}}\\[3mm]
(-1)^{\frac{k(k+1)}{2}}A_{2^{k}}^t&c_{k}J_{2^{k}}^t
\end{array}
\right),
\ee

\be
\label{i2k}
J_{2^{k+1}}= \left(
\begin{array}{cc}
0&J_{2^{k}}\\[3mm]
(-1)^{\frac{(k+1)(k+2)}{2}}J_{2^{k}}^t&0\\[3mm]
\end{array}
\right),
\ee
where $b_k, c_k \in \Cb$,
$B_{2^{k}}=b_{k}J_{2^{k}}$, $C_{2^{k}}=c_{k}J_{2^{k}}$.
We call $J_{2^{k+1}}$ multipliers.
Before proving that $A_{2^{k+1}}\in{\cal A}$, we first give the
following lemma.

{\sf Lemma 1}. $A_{2^{k+1}}$ and $J_{2^{k+1}}$ satisfy the
following relations: \be\label{ii2k}
\begin{array}{l}
J_{2^{k+1}}^tJ_{2^{k+1}}=J_{2^{k+1}}J_{2^{k+1}}^t=Id_{2^{k+1}},\\[3mm]
J_{2^{k+1}}^tJ_{2^{k+1}}^t=J_{2^{k+1}}J_{2^{k+1}}
=(-1)^{\frac{(k+1)(k+2)}{2}}Id_{2^{k+1}},
\end{array}
\ee
\be\label{ai2k}
\begin{array}{ll}
J_{2^{k+1}}^\dag =J_{2^{k+1}}^t ,~~~&
J_{2^{k+1}}^t=(-1)^{\frac{(k+1)(k+2)}{2}}J_{2^{k+1}},\\[3mm]
A_{2^{k+1}}^t=(-1)^{\frac{k(k+1)}{2}}A_{2^{k+1}},~~~&
A_{2^{k+1}}^\dag=(-1)^{\frac{k(k+1)}{2}}A^*_{2^{k+1}}.
\end{array}
\ee

{\sf Proof}. One easily checks that relations in (\ref{ii2k}) hold
for $k=1$. Suppose (\ref{ii2k}) hold for general $k$. We have
$$
\ba{rcl} J_{2^{k+1}}^tJ_{2^{k+1}}&=& \left(
\begin{array}{cc}
0&(-1)^{\frac{(k+1)(k+2)}{2}}J_{2^{k}}^t\\[3mm]
J_{2^{k}}&0\\[3mm]
\end{array}
\right) \left(
\begin{array}{cc}
0&J_{2^{k}}\\[3mm]
(-1)^{\frac{(k+1)(k+2)}{2}}J_{2^{k}}^t&0\\[3mm]
\end{array}
\right)\\[9mm]
&=&\left(
\begin{array}{cc}
(-1)^{(k+1)(k+2)}J_{2^{k}}J_{2^{k}}^t&0\\[3mm]
0&J_{2^{k}}^tJ_{2^{k}}\\[3mm]
\end{array}
\right) =Id_{2^{k+1}} \ea
$$
and
$$
\ba{rcl} J_{2^{k+1}}^tJ_{2^{k+1}}^t &=& \left(
\begin{array}{cc}
0&(-1)^{\frac{(k+1)(k+2)}{2}}J_{2^{k}}^t\\[3mm]
J_{2^{k}}&0\\[3mm]
\end{array}
\right) \left(
\begin{array}{cc}
0&(-1)^{\frac{(k+1)(k+2)}{2}}J_{2^{k}}^t\\[3mm]
J_{2^{k}}&0\\[3mm]
\end{array}
\right)\\[9mm]
&=&\left(
\begin{array}{cc}
(-1)^{\frac{(k+1)(k+2)}{2}}J_{2^{k}}J_{2^{k}}^t&0\\[3mm]
0&(-1)^{\frac{(k+1)(k+2)}{2}}J_{2^{k}}^tJ_{2^{k}}\\[3mm]
\end{array}
\right) =(-1)^{\frac{(k+1)(k+2)}{2}}Id_{2^{k+1}}. \ea
$$
Therefore the relations for $J_{2^{k+1}}^tJ_{2^{k+1}}$
and $J_{2^{k+1}}^tJ_{2^{k+1}}^t$ are valid also for $k+1$.
The cases for $J_{2^{k+1}}J_{2^{k+1}}^t$ and $J_{2^{k+1}}J_{2^{k+1}}$
can be similarly treated.

The formula $J_{2^{k+1}}^t=(-1)^{\frac{(k+1)(k+2)}{2}}J_{2^{k+1}}$
in (\ref{ai2k}) is easily deduced from (\ref{ii2k}) and the fact
$J_{2^{k+1}}^\dag = J_{2^{k+1}}^t$.

The last two formulae in (\ref{ai2k}) are easily verified for
$k=1$. If it holds for general $k$, we have then,
$$
A_{2^{k+1}}^t = \left(
\begin{array}{cc}
B_{2^{k}}^t&(-1)^{\frac{k(k+1)}{2}}A_{2^{k}}\\[3mm]
A_{2^{k}}^t&C_{2^{k}}
\end{array}
\right)
 =\left(
\begin{array}{cc}
(-1)^{\frac{k(k+1)}{2}}B_{2^{k}}&(-1)^{\frac{k(k+1)}{2}}A_{2^{k}}\\[3mm]
A_{2^{k}}^t&(-1)^{\frac{k(k+1)}{2}}C_{2^{k}}^t
\end{array}
\right) =(-1)^{\frac{k(k+1)}{2}}A_{2^{k+1}},
$$
i.e., it holds also for $k+1$. The last equality in (\ref{ai2k}) is
obtained from the conjugate of the formula above.
\hfill $\rule{2mm}{2mm}$

{\sf Lemma 2.}  The following relations can be verified
straightforwardly from Lemma 1,
\be\ba{l}\label{BC}
B_{2^{k}}^t=(-1)^{\frac{k(k+1)}{2}}B_{2^{k}},~~~
C_{2^{k}}^t=(-1)^{\frac{k(k+1)}{2}}C_{2^{k}},\\[3mm]
B_{2^{k}}^\dag =(-1)^{\frac{k(k+1)}{2}}B^*_{2^{k}},~~~
C_{2^{k}}^\dag =(-1)^{\frac{k(k+1)}{2}}C^*_{2^{k}}.
\ea
\ee
\be\label{BCI}
B_{2^{k+1}}^{-1}=\frac{1}{b_k^2}B_{2^{k+1}}^t
=\frac{1}{b_kb_k^*}B^\dag_{2^{k+1}},~~~
C_{2^{k+1}}^{-1}=\frac{1}{c_k^2}C_{2^{k+1}}^t
=\frac{1}{c_kc_k^*}C^\dag_{2^{k+1}}.
\ee
\be\label{BBCC}
\ba{l}
 B_{2^{k+1}}^tB_{2^{k+1}}=B_{2^{k+1}}B_{2^{k+1}}^t
=b_k^2 Id_{2^{k+1}},~~
C_{2^{k+1}}^tC_{2^{k+1}}=C_{2^{k+1}}C_{2^{k+1}}^t
=c_k^2 Id_{2^{k+1}},\\[3mm]
B_{2^{k+1}}^\dag B_{2^{k+1}}=B_{2^{k+1}}B^\dag_{2^{k+1}}
=b_k b_k^* Id_{2^{k+1}},~~~
C_{2^{k+1}}^\dag C_{2^{k+1}}=C_{2^{k+1}}C_{2^{k+1}}^\dag
=c_k c_k^* Id_{2^{k+1}}.
\ea
\ee

\medskip
For any $A_{2^{k+1}}\in \cal A$, $k\geq 2$, we define
\be
\ba{rcl}
||A_{2^{k+1}}||&=:&b_kb_k+c_kc_k+||A_{2^k}||,\\[3mm]
[A_{2^{k+1}}]&=:&(-1)^{k(k+1)/2}b_kc_k-[A_{2^k}].
\ea
\ee

{\sf Lemma 3}. For any $k\geq 2$, we have, \be\label{lemma3}
\ba{rcl} (A_{2^{k+1}}J_{2^{k+1}})(J_{2^{k+1}}A_{2^{k+1}})^t&=&
(A_{2^{k+1}}J_{2^{k+1}})^t(J_{2^{k+1}}A_{2^{k+1}})\\[3mm]
&=&((-1)^{\frac{k(k+1)}{2}}b_kc_k-[A_{2^{k}}])Id_{2^{k+1}}
=[A_{2^{k+1}}] Id_{2^{k+1}},\\[4mm]
(A_{2^{k+1}}^\ast J_{2^{k+1}})(J_{2^{k+1}}A_{2^{k+1}}^\ast)^t&=&
(A_{2^{k+1}}^\ast J_{2^{k+1}})^t(J_{2^{k+1}}A_{2^{k+1}}^\ast)
=[A_{2^{k+1}}]^\ast Id_{2^{k+1}}.
\ea
\ee

{\sf Proof}. One can verify that Lemma 3 holds for $k=2$. Suppose
it is valid for $k$, we have
$$
\ba{l}
(A_{2^{k+1}}J_{2^{k+1}})(J_{2^{k+1}}A_{2^{k+1}})^t\\[3mm]
=\left(
\begin{array}{cc}
(-1)^{\frac{(k+1)(k+2)}{2}}A_{2^{k}}J_{2^{k}}^t&B_{2^{k}}J_{2^{k}}\\[3mm]
(-1)^{\frac{(k+1)(k+2)}{2}}C_{2^{k}}^tJ_{2^{k}}^t&
(-1)^{\frac{k(k+1)}{2}}A_{2^{k}}^tJ_{2^{k}}
\end{array}
\right) \left(
\begin{array}{cc}
(-1)^{\frac{k(k+1)}{2}}J_{2^{k}}A_{2^{k}}^t&J_{2^{k}}C_{2^{k}}^t\\[3mm]
(-1)^{\frac{(k+1)(k+2)}{2}}J_{2^{k}}^tB_{2^{k}}&
(-1)^{\frac{(k+1)(k+2)}{2}}J_{2^{k}}^tA_{2^{k}}
\end{array}
\right)^t\\[9mm]
=\left(
\begin{array}{cc}
e_{11}& e_{12}\\[3mm]
e_{21}& e_{22}
\end{array}
\right), \ea
$$
where
$$
\ba{rcl}
e_{11}&=&(-1)^{\frac{(k+1)(k+2)+k(k+1)}{2}}A_{2^{k}}J_{2^{k}}^tA_{2^{k}}J_{2^{k}}^t
+(-1)^{\frac{k(k+1)}{2}}b_kc_k Id_{2^{k}}\\[3mm]
&=&(-1)^{\frac{(k+1)(k+2)+k(k-1)}{2}}(A_{2^{k}}J_{2^{k}}^t)(J_{2^{k}}A_{2^{k}}^t)^t
+(-1)^{\frac{k(k+1)}{2}}b_kc_k Id_{2^{k}}\\[3mm]
&=&((-1)^{\frac{k(k+1)}{2}}b_kc_k-[A_{2^{k}}])Id_{2^{k}},\\[3mm]
e_{12}&=&b_k A_{2^{k}}J_{2^{k}}^t+
(-1)^{\frac{(k+1)(k+2)+k(k+1)}{2}}b_kA_{2^{k}}^tJ_{2^{k}}\\[3mm]
&=& b_kA_{2^{k}}J_{2^{k}}^t(1+(-1)^{\frac{(k+1)(k+2)+k(k-1)}{2}})=0,\\[3mm]
e_{21}&=&(-1)^{\frac{(k+1)(k+2)}{2}}c_k A_{2^{k}}J_{2^{k}}^t
+(-1)^{\frac{k(k+1)}{2}}c_k A_{2^{k}}^tJ_{2^{k}}=0,\\[3mm]
e_{22}&=& (-1)^{\frac{k(k+1)}{2}}b_k c_kId_{2^{k}}+
(-1)^{\frac{(k+1)(k+2)+k(k+1)}{2}}A_{2^{k}}^tJ_{2^{k}}A_{2^{k}}^tJ_{2^{k}}\\[3mm]
&=&(-1)^{\frac{k(k+1)}{2}}b_k c_k Id_{2^{k}}+
(-1)^{\frac{(k+1)(k+2)+k(k-1)}{2}}(A_{2^{k}}J_{2^{k}})(J_{2^{k}}A_{2^{k}})^t\\[3mm]
&=&((-1)^{\frac{k(k+1)}{2}}b_kc_k-[A_{2^{k}}])Id_{2^{k}}. \ea
$$
Hence
$$
(A_{2^{k+1}}J_{2^{k+1}})(J_{2^{k+1}}A_{2^{k+1}})^t
=((-1)^{\frac{k(k+1)}{2}}b_kc_k-[A_{2^{k}}])Id_{2^{k+1}}=[A_{2^{k+1}}])Id_{2^{k+1}}.
$$
Similar calculations apply to
$(A_{2^{k+1}}J_{2^{k+1}})^t(J_{2^{k+1}}A_{2^{k+1}})$. Therefore the Lemma holds
for $k+1$. The last equation can be deduced from the first one.

{\sf Theorem 2}. $A_{2^{k}}$ satisfies the following relation: \be
\label{thm2}
|A_{2^{k+1}}A_{2^{k+1}}^\dag|=([A_{2^{k+1}}][A_{2^{k+1}}]^*)^{2^k}
=[((-1)^{\frac{k(k+1)}{2}}b_kc_k-[A_{2^{k}}])
((-1)^{\frac{k(k+1)}{2}}b^*_kc^*_k-[A_{2^{k}}]^*)]^{2^k}. \ee

{\sf Proof}. By using Lemma 1-3, we have
$$
\ba{rcl}
|A_{2^{k+1}}|
&=&\left|
\left(
\begin{array}{cc}
B_{2^{k}}& A_{2^{k}}\\[3mm]
(-1)^{\frac{k(k+1)}{2}}A_{2^{k}}^t&C_{2^{k}}^t
\end{array}
\right)
\right|\\[9mm]
&=&\left|
\left(
\begin{array}{cc}
Id_{2^{k}}&-A_{2^{k}}(C_{2^{k}}^t)^{-1}\\[3mm]
0&Id_{2^{k}}
\end{array}
\right)
\left(
\begin{array}{cc}
B_{2^{k}}& A_{2^{k}}\\[3mm]
(-1)^{\frac{k(k+1)}{2}}A_{2^{k}}^t&C_{2^{k}}^t
\end{array}
\right)
\right|\\[9mm]
&=&\left|
\left(
\begin{array}{cc}
B_{2^{k}}-(-1)^{\frac{k(k+1)}{2}}A_{2^{k}}(C_{2^{k}}^t)^{-1}A_{2^{k}}^t& 0\\[3mm]
(-1)^{\frac{k(k+1)}{2}}A_{2^{k}}^t&C_{2^{k}}^t
\end{array}
\right)
\right|\\[9mm]
&=&|b_kc_k J_{2^{k}}J_{2^{k}}^t-(-1)^{\frac{k(k+1)}{2}}\frac{1}{c_k^2}
A_{2^{k}}C_{2^{k}}A_{2^{k}}^tC_{2^{k}}^t|\\[4mm]
&=&|b_kc_k Id_{2^{k}}-(-1)^{\frac{k(k+1)}{2}}
(A_{2^{k}}J_{2^{k}})(J_{2^{k}}A_{2^{k}})^t|\\[4mm]
&=&|(-1)^{\frac{k(k+1)}{2}}b_kc_k Id_{2^{k}}-
[A_{2^{k}}]Id_{2^{k}}|
=((-1)^{\frac{k(k+1)}{2}}b_kc_k-[A_{2^{k}}])^{2^k}.
\ea
$$
Therefore
$$
|A_{2^{k+1}}A_{2^{k+1}}^\dag|=([A_{2^{k+1}}][A_{2^{k+1}}]^\ast)^{2^k}.
$$
\hfill $\rule{2mm}{2mm}$

{\sf Lemma 4}. $A_{2^{k+1}}$ and $J_{2^{k+1}}$ satisfy the following relations:
$$
\ba{l}
(A_{2^{k+1}}J_{2^{k+1}})(J_{2^{k+1}}A_{2^{k+1}})^\dag
+(J_{2^{k+1}}A_{2^{k+1}}^\ast)(J_{2^{k+1}}A_{2^{k+1}})^t\\[3mm]
~~~~~~~~=A_{2^{k+1}}A_{2^{k+1}}^\dag+J_{2^{k+1}}A_{2^{k+1}}^\ast
A_{2^{k+1}}^tJ_{2^{k+1}}^t
=||A_{2^{k+1}}||Id_{2^{k+1}},\\[3mm]
(A_{2^{k+1}}J_{2^{k+1}})^t(A_{2^{k+1}}^\ast J_{2^{k+1}})
+(J_{2^{k+1}}A_{2^{k+1}})^\dag(J_{2^{k+1}}A_{2^{k+1}})\\[3mm]
~~~~~~~~=A_{2^{k+1}}^\dag A_{2^{k+1}}+J_{2^{k+1}}^tA_{2^{k+1}}^tA_{2^{k+1}}^\ast J_{2^{k+1}}
=||A_{2^{k+1}}||Id_{2^{k+1}}.
\ea
$$

{\sf Proof}. It can be verified that the first formula
holds for $k=2$, if it holds for $k$, we have
$$
\ba{l} (A_{2^{k+1}}J_{2^{k+1}})(A_{2^{k+1}}J_{2^{k+1}})^\dag
+(J_{2^{k+1}}A^*_{2^{k+1}})(J_{2^{k+1}}A_{2^{k+1}})^t\\[3mm]
=\left(
\begin{array}{cc}
(-1)^{\frac{(k+1)(k+2)}{2}}A_{2^{k}}J_{2^{k}}^t&B_{2^{k}}J_{2^{k}}\\[3mm]
(-1)^{\frac{(k+1)(k+2)}{2}}C_{2^{k}}^tJ_{2^{k}}^t&
(-1)^{\frac{k(k+1)}{2}}A_{2^{k}}^tJ_{2^{k}}
\end{array}
\right) \left(
\begin{array}{cc}
(-1)^{\frac{(k+1)(k+2)}{2}}J_{2^{k}}A_{2^{k}}^\dag&
(-1)^{\frac{(k+1)(k+2)}{2}}J_{2^{k}}C^*_{2^{k}}\\[3mm]
J_{2^{k}}^tB_{2^{k}}^\dag&
(-1)^{\frac{k(k+1)}{2}}J^t_{2^{k}}A_{2^{k}}^*
\end{array}
\right)\\[9mm]
+\left(
\begin{array}{cc}
(-1)^{\frac{k(k+1)}{2}}J_{2^{k}}A_{2^{k}}^\dag&J_{2^{k}}C_{2^{k}}^\dag\\[3mm]
(-1)^{\frac{(k+1)(k+2)}{2}}J_{2^{k}}^tB_{2^{k}}^*&
(-1)^{\frac{(k+1)(k+2)}{2}}J_{2^{k}}^tA_{2^{k}}^*
\end{array}
\right) \left(
\begin{array}{cc}
(-1)^{\frac{k(k+1)}{2}}A_{2^{k}}J_{2^{k}}^t&
(-1)^{\frac{(k+1)(k+2)}{2}}B_{2^{k}}^tJ_{2^{k}}\\[3mm]
C_{2^{k}}J_{2^{k}}^t&
(-1)^{\frac{(k+1)(k+2)}{2}}A_{2^{k}}^tJ_{2^{k}}
\end{array}
\right)\\[9mm]
=\left(
\begin{array}{cc}
f_{11}& f_{12}\\[3mm]
f_{21}& f_{22}
\end{array}
\right), \ea
$$
where, by using Lemma 1 and 2,
$$
\ba{rcl}
f_{11}=f_{22}&=& A_{2^{k}}A_{2^{k}}^\dag+J_{2^{k}}A_{2^{k}}^\dag
A_{2^{k}}J_{2^{k}}^t+BB^\dag +J_{2^{k}}C^\dag CJ_{2^{k}}^t \\[3mm]
&=&
A_{2^{k}}A_{2^{k}}^\dag+J_{2^{k}}(-1)^{\frac{k(k+1)}{2}}A_{2^{k}}^*
(-1)^{\frac{k(k+1)}{2}}A^t_{2^{k}}J_{2^{k}}^t+(b_kb_k^*+c_kc_k^*)Id_{2^{k}}\\[3mm]
&=&(||A_{2^{k}}||+b_kb_k^*+c_kc_k^*)Id_{2^{k}}=||A_{2^{k+1}}||Id_{2^{k}},
\ea
$$
$$
\ba{rcl}
f_{12}&=&A_{2^{k}}C^*_{2^{k}}+(-1)^{\frac{k(k+1)}{2}}B_{2^{k}}A^*_{2^{k}}
+(-1)^{\frac{k(k+1)+(k+1)(k+2)}{2}}b_kJ_{2^{k}}A_{2^{k}}^\dag
+(-1)^{\frac{(k+1)(k+2)}{2}}c^*_kA_{2^{k}}^tJ_{2^{k}}\\[3mm]
&=&(-1)^{\frac{k(k+1)}{2}}(B_{2^{k}}A^*_{2^{k}}
+(-1)^{\frac{k(k-1)+(k+1)(k+2)}{2}}B_{2^{k}}A^*_{2^{k}})\\[3mm]
&&+A_{2^{k}}C^*_{2^{k}}
+(-1)^{\frac{k(k-1)+(k+1)(k+2)}{2}}A_{2^{k}}C^*_{2^{k}}=0, \ea
$$
$$
\ba{rcl}
f_{21}&=&C_{2^{k}}^tA^\dag_{2^{k}}+(-1)^{\frac{k(k+1)}{2}}A_{2^{k}}^tB_{2^{k}}^\dag
+(-1)^{\frac{k(k+1)+(k+1)(k+2)}{2}}b^*_kA_{2^{k}}J_{2^{k}}^t
+(-1)^{\frac{(k+1)(k+2)}{2}}c_kJ_{2^{k}}^tA^*_{2^{k}}\\[3mm]
&=&(-1)^{\frac{k(k+1)}{2}}(b_kA_{2^{k}}^tJ_{2^{k}}^t
+(-1)^{\frac{k(k-1)+(k+1)(k+2)}{2}}b_kA_{2^{k}}^tJ_{2^{k}}^t)\\[3mm]
&&+c^*_k J_{2^{k}}^tA_{2^{k}}^t
+(-1)^{\frac{k(k-1)+(k+1)(k+2)}{2}}c_k^* J_{2^{k}}^tA_{2^{k}}^t
=0. \ea
$$
Hence the first formula holds also for $k+1$.
The second formula can be verified similarly.
\hfill $\rule{2mm}{2mm}$

{\sf Lemma 5}. Matrices $B_{2^k}$, $A_{2^k}$ and $C_{2^k}$
satisfy the following relations:
\be\label{fa2k}
((-1)^{k(k+1)\over 2} B_{2^k}A_{2^k}^\ast+A_{2^k}C_{2^k}^\ast)
((-1)^{k(k+1)\over 2}A_{2^k}^\ast B_{2^k}+C_{2^k}^\ast A_{2^k})^t
= F(A_{2^{k+1}})Id_{2^k},
\ee
where
\begin{equation}\label{F}
F(A_{2^{k+1}}) =c_k^{\ast 2}[A_{2^k}]+b_k^2[A_{2^k}]^{\ast}
+(-1)^{k(k+1)\over 2} b_kc_k^{\ast} \| A_{2^k} \|.
\end{equation}

{\sf Proof}. By using Lemma 3 and 4, we have
$$
\begin{array}{l}
((-1)^{k(k+1)\over 2} B_{2^k}A_{2^k}^\ast+A_{2^k}C_{2^k}^\ast)
((-1)^{k(k+1)\over 2}A_{2^k}^\ast B_{2^k}+C_{2^k}^\ast A_{2^k})^t\\[3mm]
=b_k^2(J_{2^k}A_{2^k}^{\ast})(A_{2^k}^{\ast}J_{2^k})^t
+c_k^{\ast 2}(A_{2^k}J_{2^k})(J_{2^k}A_{2^k})^t\\[3mm]
~~~+(-1)^{k(k+1)\over 2}b_k c_k^{\ast}[(A_{2^k}^{\ast}J_{2^k})(A_{2^k}^{\ast}J_{2^k})^t
+(J_{2^k}A_{2^k}^{\ast})(J_{2^k}A_{2^k})^t]\\[3mm]
=(c_k^{\ast 2}[A_{2^k}]+b_k^2[A_{2^k}]^{\ast}
+(-1)^{k(k+1)\over 2} b_kc_k^{\ast} \| A_{2^k} \|) Id_{2^k}
=F(A_{2^{k+1}})Id_{2^k}.
\end{array}
$$

{\sf Lemma 6}. $A_{2^k}$ and $J_{2^k}$ satisfy the following relation:
\be\label{fa2kp}
\| A_{2^k} \| J_{2^k}A_{2^k}^\ast
A_{2^k}^tJ_{2^k}^t = [ A_{2^k} ][ A_{2^k} ]^\ast Id_{2^k} +
J_{2^k}A_{2^k}^\ast A_{2^k}^tA_{2^k}^\ast A_{2^k}^tJ_{2^k}^t. \ee

{\sf Proof}. From (\ref{fa2k}) we have the following relation:
$$
\begin{array}{l}
F(A_{2^{k+1}})J_{2^k}A_{2^k}^\ast A_{2^k}^tJ_{2^k}^t\\[4mm]
=((-1)^{k(k+1)\over 2}B_{2^k}A_{2^k}^\ast +A_{2^k}C_{2^k}^\ast )(-1)^{k(k+1)\over 2} b_k
(A_{2^k}^\ast J_{2^k})^tJ_{2^k}A_{2^k}^\ast A_{2^k}^tJ_{2^k}^t\\[4mm]
    \quad+ ((-1)^{k(k+1)\over 2} B_{2^k}A_{2^k}^\ast +A_{2^k}C_{2^k}^\ast )c_k^\ast
    (J_{2^k}A_{2^k})^tJ_{2^k}A_{2^k}^\ast A_{2^k}^tJ_{2^k}^t\\[4mm]
=(-1)^{k(k+1)\over 2} b_k((-1)^{k(k+1)\over 2} B_{2^k}A_{2^k}^\ast +A_{2^k}C_{2^k}^\ast )
  [(A_{2^k}^\ast J_{2^k})^t(J_{2^k}A_{2^k}^\ast) ] A_{2^k}^tJ_{2^k}^t\\[4mm]
    \quad+ c_k^\ast [(-1)^{k(k+1)\over 2} B_{2^k}A_{2^k}^\ast
    (J_{2^k}A_{2^k})^tJ_{2^k}A_{2^k}^\ast A_{2^k}^tJ_{2^k}^t
 +A_{2^k}C_{2^k}^\ast (J_{2^k}A_{2^k})^tJ_{2^k}A_{2^k}^\ast A_{2^k}^tJ_{2^k}^t ]\\[4mm]
=(-1)^{k(k+1)\over 2} b_k((-1)^{k(k+1)\over 2} b_kJ_{2^k}A_{2^k}^\ast
[ A_{2^k} ] A_{2^k}^tJ_{2^k}^t + c_k^\ast A_{2^k}J_{2^k} [A_{2^k}]^\ast A_{2^k}^tJ_{2^k}^t )\\[4mm]
    \quad + c_k^\ast [ (-1)^{k(k+1)\over 2} b_kJ_{2^k}A_{2^k}^\ast
    A_{2^k}^tA_{2^k}^\ast A_{2^k}^tJ_{2^k}^t
          +c_k^\ast  [A_{2^k}] J_{2^k}A_{2^k}^\ast A_{2^k}^tJ_{2^k}^t ]\\[4mm]
=b_k^2[ A_{2^k} ]^\ast J_{2^k}A_{2^k}^\ast  A_{2^k}^tJ_{2^k}^t + (-1)^{k(k+1)\over 2}
b_kc_k^\ast  [ A_{2^k} ][ A_{2^k} ]^\ast Id_{2^k})\\[4mm]
    \quad +(-1)^{k(k+1)\over 2} c_k^\ast
    b_kJ_{2^k}A_{2^k}^\ast A_{2^k}^tA_{2^k}^\ast A_{2^k}^tJ_{2^k}^t
+c_2^2 [A_{2^k}] J_{2^k}A_{2^k}A_{2^k}^tJ_{2^k}^t ]\\[4mm]
=(b_k^2 + c_k^2)[ A_{2^k} ]J_{2^k}A_{2^k}
A_{2^k}^tJ_{2^k}^t + (-1)^{k(k+1)\over 2}
b_kc_k ([ A_{2^k} ] ^2 Id_{2^k}\\[4mm]
\quad +(-1)^{k(k+1)\over 2} c_k b_kJ_{2^k}A_{2^k}A_{2^k}^tA_{2^k}A_{2^k}^tJ_{2^k}^t).
\end{array}
$$

Using (\ref{F}) we have
$$
\| A_{2^k} \| J_{2^k}A_{2^k}^\ast A_{2^k}^tJ_{2^k}^t
= [ A_{2^k} ][ A_{2^k} ]^\ast Id_{2^k}
+ J_{2^k}A_{2^k}^\ast A_{2^k}^tA_{2^k}^\ast A_{2^k}^tJ_{2^k}^t.
$$
\hfill $\rule{2mm}{2mm}$

{\sf Theorem 3}. The eigenvalue polynom of
$A_{2^{k+1}}A_{2^{k+1}}^\dag$ satisfies the following relations:
\be
\label{thm4}
\ba{l} |A_{2^{k+1}}A_{2^{k+1}}^\dag-\lambda
Id_{2^{k+1}} |
=(\lambda^2-||A_{2^{k+1}}||\lambda+[A_{2^{k+1}}][A_{2^{k+1}}]^*)^{2^k},\\[3mm]
|A_{2^{k+1}}^\dag A_{2^{k+1}}-\lambda Id_{2^{k+1}} |
=(\lambda^2-||A_{2^{k+1}}||\lambda+[A_{2^{k+1}}][A_{2^{k+1}}]^*)^{2^k}.
\ea
\ee

{\sf Proof}.
Let
$$
\Lambda_k=-[(c_kc_k^\ast-\lambda) Id_{2^k} + A_{2^k}^tA_{2^k}^\ast] [(-1)^{k(k+1)\over 2}
B_{2^k}A_{2^k}^\ast+A_{2^k}C_{2^k}^\ast]^{-1}.
$$

$$
\begin{array}{l}
\left|A_{2^{k+1}}A_{2^{k+1}}^\dag - \lambda Id_{2^{k+1}} \right|
= \left|
\left(
\begin{array}{cc}
(-1)^{k(k+1)\over 2} B_{2^k}A_{2^k}^\ast+A_{2^k}C_{2^k}^\ast&(b_kb_k^\ast-\lambda)Id_{2^k}
+ A_{2^k}A_{2^k}^\dag  \\[3mm]
(c_kc_k^\ast-\lambda) Id_{2^k} + A_{2^k}^tA_{2^k}^\ast&(-1)^{k(k+1)\over 2} A_{2^k}^tB_{2^k}^\dag
+C_{2^k}^tA_{2^k}^\dag  \\[3mm]
\end{array}
\right)
\right|\\[9mm]
=\left|
\left(
\begin{array}{cc}
Id_{2^k}&0\\[3mm]
\Lambda_k&Id_{2^k}\\[3mm]
\end{array}\right)
\left(
\begin{array}{cc}
(-1)^{k(k+1)\over 2} B_{2^k}A_{2^k}^\ast+A_{2^k}C_{2^k}^\ast&(b_kb_k^\ast-\lambda)Id_{2^k}
+ A_{2^k}A_{2^k}^\dag  \\[3mm]
(c_kc_k^\ast-\lambda) Id_{2^k} + A_{2^k}^tA_{2^k}^\ast&(-1)^{k(k+1)\over 2} A_{2^k}^tB_{2^k}^\dag
+C_{2^k}^tA_{2^k}^\dag  \\[3mm]
\end{array}
\right)\right|\\[9mm]
=\left|
\left(\begin{array}{cc}
(-1)^{k(k+1)\over 2} B_{2^k}A_{2^k}^\ast+A_{2^k}C_{2^k}^\ast&(b_kb_k^\ast-\lambda)Id_{2^k}
+ A_{2^k}A_{2^k}^\dag  \\[3mm]
0&-\Lambda_k[(b_kb_k^\ast-\lambda)Id_{2^k} + A_{2^k}A_{2^k}^\dag]
           +(-1)^{k(k+1)\over 2} A_{2^k}^tB_{2^k}^\dag+C_{2^k}^tA^\dag_{2^k}
\end{array}
\right)
\right|\\[9mm]
=\left|I+II \right|,
\end{array}
$$
where
$$
\ba{rcl}
I&=&((-1)^{k(k+1)\over 2} B_{2^k}A_{2^k}^\ast+A_{2^k}C_{2^k}^\ast)
((-1)^{k(k+1)\over 2} B_{2^k}^\ast A_{2^k}+A_{2^k}^\ast C_{2^k})^t\\[3mm]
&=&(-1)^{k(k+1)\over 2}b_kc_k[A_{2^k}]^\ast Id_{2^k}
+(-1)^{k(k+1)\over 2}b_k^\ast c_k^\ast [A_{2^k}]Id_{2^k}
+b_kb_k^\ast J_{2^k}A_{2^k}^\ast A_{2^k}^tJ_{2^k}^t
+c_kc_k^\ast A_{2^k}A_{2^k}^\dag
\ea
$$
and, by using Lemma 5,
$$
\ba{rcl}
II&=&-((-1)^{k(k+1)\over 2} B_{2^k}A_{2^k}^\ast +A_{2^k}C_{2^k}^\ast )
\Lambda_k[(b_kb_k^\ast -\lambda)Id_{2^k} + A_{2^k}A_{2^k}^\dag]\\[3mm]
&=&[(c_kc_k^\ast -\lambda)((-1)^{k(k+1)\over 2}B_{2^k}A_{2^k}^\ast +A_{2^k}C_{2^k}^\ast)
+((-1)^{k(k+1)\over 2}B_{2^k}A_{2^k}^\ast +A_{2^k}C_{2^k}^\ast )A_{2^k}^tA_{2^k}^\ast]\\[3mm]
&&[(b_kb_k^\ast -\lambda)((-1)^{k(k+1)\over 2}B_{2^k}A_{2^k}^\ast +A_{2^k}C_{2^k}^\ast)^{-1}
+((-1)^{k(k+1)\over 2}B_{2^k}A_{2^k}^\ast +A_{2^k}C_{2^k}^\ast)^{-1}A_{2^k}A_{2^k}^\dag]\\[3mm]
&=&(b_kb_k^\ast -\lambda)(c_kc_k^\ast -\lambda)Id_{2^k}+(b_kb_k^\ast -\lambda)
((-1)^{k(k+1)\over 2}B_{2^k}A_{2^k}^\ast \\[3mm]
&&+A_{2^k}C_{2^k}^\ast )A_{2^k}^tA_{2^k}^\ast
((-1)^{k(k+1)\over 2}B_{2^k}A_{2^k}^\ast +A_{2^k}C_{2^k}^\ast )^{-1}
+(c_kc_k^\ast -\lambda)A_{2^k}A_{2^k}^\dag\\[3mm]
&&+((-1)^{k(k+1)\over 2}B_{2^k}A_{2^k}^\ast +A_{2^k}C_{2^k}^\ast)A_{2^k}^tA_{2^k}^\ast
((-1)^{k(k+1)\over 2}B_{2^k}A_{2^k}^\ast +A_{2^k}C_{2^k}^\ast)^{-1}A_{2^k}A_{2^k}^\dag\\[3mm]
&=&(b_kb_k^\ast -\lambda)(c_kc_k^\ast -\lambda)Id_{2^k}+
(c_kc_k^\ast -\lambda)A_{2^k}A_{2^k}^\dag
+\frac{b_kb_k^\ast -\lambda}{F(A_{2^{k+1}})}III
+\frac{1}{F(A_{2^{k+1}})}III A_{2^k}A_{2^k}^\dag,
\ea
$$
where
$$
\ba{rcl}
III&=&((-1)^{k(k+1)\over 2}B_{2^k}A_{2^k}^\ast
+A_{2^k}C_{2^k}^\ast )A_{2^k}^tA_{2^k}^\ast
((-1)^{k(k+1)\over 2}A_{2^k}^\ast B_{2^k}+C_{2^k}^\ast A_{2^k})^t\\[3mm]
&=&((-1)^{k(k+1)\over 2}B_{2^k}A_{2^k}^\ast +A_{2^k}C_{2^k}^\ast )
A_{2^k}^tJ_{2^k}^tJ_{2^k}A_{2^k}^\ast
((-1)^{k(k+1)\over 2}A_{2^k}^\ast B_{2^k}+C_{2^k}^\ast A_{2^k})^t\\[3mm]
&=&[(-1)^{k(k+1)\over 2}b_k (J_{2^k}A_{2^k}^\ast )(J_{2^k}A_{2^k})^t
+c_k^\ast (A_{2^k}J_{2^k})(J_{2^k}A_{2^k})^t]\\[3mm]
&&[(-1)^{k(k+1)\over 2}b_k (J_{2^k}A_{2^k}^\ast )(A_{2^k}^\ast J_{2^k})^t
+c_k^\ast (J_{2^k}A_{2^k}^\ast )(J_{2^k}A_{2^k})^t]\\[3mm]
&=&[(-1)^{k(k+1)\over 2}b_k (J_{2^k}A_{2^k}^\ast )(J_{2^k}A_{2^k})^t
+c_k^\ast [A_{2^k}]Id_{2^k}]\cdot\\[3mm]
&&~~[(-1)^{k(k+1)\over 2}b_k [A_{2^k}]^\ast Id_{2^k}
+c_k^\ast (J_{2^k}A_{2^k}^\ast )(J_{2^k}A_{2^k})^t]\\[3mm]
&=&(b_k^2[A_{2^k}]^\ast+c_k^{\ast 2}[A_{2^k}])J_{2^k}A_{2^k}^\ast A_{2^k}^tJ_{2^k}^t
+(-1)^{k(k+1)\over 2}b_kc_k^\ast J_{2^k}A_{2^k}^\ast
A_{2^k}^tA_{2^k}^\ast A_{2^k}^tJ_{2^k}^t\\[3mm]
&&+(-1)^{k(k+1)\over 2}b_kc_k^\ast [A_{2^k}][A_{2^k}]^\ast Id_{2^k}.\\[3mm]
\ea
$$
From Lemma 6, we get
$$
\ba{rcl}
III&=&(b_k^2[A_{2^k}]^\ast+c_k^{\ast 2}[A_{2^k}])J_{2^k}A_{2^k}^\ast A_{2^k}^tJ_{2^k}^t
+(-1)^{k(k+1)\over 2}b_kc_k^\ast ||A_{2^k}||J_{2^k}A_{2^k}^\ast A_{2^k}^tJ_{2^k}^t\\[3mm]
&=&F(A_{2^{k+1}})J_{2^k}A_{2^k}^\ast A_{2^k}^tJ_{2^k}^t.
\ea
$$

From Lemma 3 we also have
$$
\ba{rcl}
III A_{2^k}A_{2^k}^\dag&=&III A_{2^k}J_{2^k} J_{2^k}^tA_{2^k}^\dag\\[3mm]
&=&F(A_{2^{k+1}})J_{2^k}A_{2^k}^\ast(J_{2^k}A_{2^k})^t(A_{2^k}J_{2^k})
(A_{2^k}^\ast J_{2^k})^t
=F(A_{2^{k+1}})[A_{2^k}][A_{2^k}]^\ast Id_{2^k}.
\ea
$$

Therefore,
$$
\ba{l}
\left|A_{2^{k+1}}A_{2^{k+1}}^\dag - \lambda Id_{2^{k+1}} \right|
=|I+II|\\[3mm]
=|-\lambda^2 Id_{2^k}+\lambda(b_kb_k^\ast+c_kc_k^\ast+||A_{2^k}||)Id_{2^k}
-(b_kb_k^\ast c_kc_k^\ast-(-1)^{k(k+1)\over 2}b_k^\ast c_k^\ast[A_{2^k}]\\[3mm]
~~~-(-1)^{k(k+1)\over 2}b_kc_k[A_{2^k}]^\ast+[A_{2^k}][A_{2^k}]^\ast)Id_{2^k}|\\[3mm]
=(\lambda^2-||A_{2^{k+1}}||\lambda+[A_{2^{k+1}}][A_{2^{k+1}}]^\ast)^{2^k},
\ea
$$
where the first formula in Lemma 4 is used. The second formula in Theorem 3
is obtained from the fact that $A_{2^{k+1}}A_{2^{k+1}}^\dag$  and
$A_{2^{k+1}}^\dag A_{2^{k+1}}$ have the same eigenvalue set.
\hfill $\rule{2mm}{2mm}$

From Theorem 2 and 3 the states given by (\ref{a2k}) are $d$-computable.
In terms of (\ref{GC}) the generalized concurrence for these states is given by
$$
d_{2^{k+1}}=2^{k+1}\vert[A_{2^{k+1}}]\vert=2^{k+1}\vert b_kc_k+
b_{k-1}c_{k-1}+...+b_1c_1+ad+c^2\vert.
$$

Let $p_{2^{k+1}}$ be a symmetric anti-diagonal $2^{2k+2}\times 2^{2k+2}$ matrix with
all the anti-diagonal elements $1$ except for those
at rows $2^{k+1}-1 + s(2^{k+2}-2)$, $2^{k+1} + s(2^{k+2}-2)$,
$2^{k+2}-1 + s(2^{k+2}-2)$, $2^{k+2} + s(2^{k+2}-2)$,
$s=0,...,2^{k+1}-1$, which are $-1$. $d_{2^{k+1}}$ can then be written as
\be\label{dkp}
d_{2^{k+1}}=|\langle \psi_{2^{k+1}} |p_{2^{k+1}}\psi_{2^{k+1}}^{*} \rangle |,
\ee
where
\be\label{psi2k1}
\vert\psi_{2^{k+1}}\rangle=\sum_{i,j=1}^{2^{k+1}} (A_{2^{k+1}})_{ij}\,e_i\otimes e_j.
\ee

Let $\Phi$ denote the set of pure states with the form (\ref{psi2k1}).
For mixed states with
density matrices such that their decompositions are of the form
\be\label{rho12k1}
\rho_{2^{2k+2}} = \sum_{i=1}^M p_i |\psi_i \rangle \langle \psi_i|,~~~~\sum_{i=1}^M
p_i =1,~~~~|\psi_i\rangle\in\Phi,
\ee
their entanglement of formations,
by using a similar calculation in obtaining formula (\ref{drho}) \cite{fjlw},
are then given by $E(d_{2^{k+1}}(\rho_{2^{2k+2}}))$, where
\be\label{d2k1}
d_{2^{k+1}}(\rho_{2^{2k+2}})=\Omega_1 - \sum_{i=2}^{2^{2k+2}}\Omega_i,
\ee
and $\Omega_i$, in decreasing order, are the
the square roots of the eigenvalues of the
matrix $\rho_{2^{2k+2}} p_{2^{k+1}}{\rho_{2^{2k+2}}^\ast}p_{2^{k+1}}$.
Here again due to the form of the so constructed matrix $A_{2^{k+1}}$ in (\ref{a2k}),
once $\rho$ has a decomposition with all $|\psi_i\rangle\in\Phi$, all
other decompositions of $|\psi_i^\prime\rangle$
also satisfy $|\psi_i^\prime\rangle\in\Phi$.
Therefore from high dimensional $d$-computable states $A_{2^{k+1}}$,
$2\leq k\leq N$, the entanglement of formation for a class of density matrices
whose decompositions lie in these $d$-computable quantum states can be obtained
analytically.

\section{Remarks and conclusions}

Besides the $d$-computable states constructed above,
from (\ref{ha4}) we can also construct another class of high dimensional
$d$-computable states given by $2^{k+1}\times 2^{k+1}$ matrices $A_{2^{k+1}}$,
$2\leq  k\in\Nb$,
\be\label{newa}
A_{2^{k+1}}= \left(
\begin{array}{cc}
B_k&A_k\\[3mm]
-A_k^t&C_k\\[3mm]
\end{array}
\right)
\equiv\left(
\begin{array}{cc}
b_{k}I_{2^{k}}&A_{2^{k}}\\[3mm]
-A_{2^{k}}^t&c_{k}I_{2^{k}}
\end{array}
\right),
\ee
where $b_k,~c_k\in\Cb$, $I_4=J_4$,
\be
I_{2^{k+1}}= \left(
\begin{array}{cc}
0&I_{2^{k}}\\[3mm]
-I_{2^{k}}&0\\[3mm]
\end{array}
\right)
\ee
for $k+2~ mode~4 =0$,
\be
I_{2^{k+1}}= \left(
\begin{array}{cc}
0&I_{2^{k}}\\[3mm]
I_{2^{k}}&0
\end{array}
\right)
\ee
for $k+2~ mode~4 =1$,
\be
I_{2^{k+1}}= \left(
\begin{array}{cccc}
0&0&0&I_{2^{k-1}}\\[3mm]
0&0&-I_{2^{k-1}}&0\\[3mm]
0&I_{2^{k-1}}&0&0\\[3mm]
-I_{2^{k-1}}&0&0&0
\end{array}
\right)
\ee
for $k+2~ mode~4 =2$, and
\be
I_{2^{k+1}}= \left(
\begin{array}{cccccccc}
0&0&0&0&0&0&0&I_{2^{k-2}}\\[3mm]
0&0&0&0&0&0&-I_{2^{k-2}}&0\\[3mm]
0&0&0&0&0&-I_{2^{k-2}}&0&0\\[3mm]
0&0&0&0&I_{2^{k-2}}&0&0&0\\[3mm]
0&0&0&-I_{2^{k-2}}&0&0&0&0\\[3mm]
0&0&I_{2^{k-2}}&0&0&0&0&0\\[3mm]
0&I_{2^{k-2}}&0&0&0&0&0&0\\[3mm]
-I_{2^{k-2}}&0&0&0&0&0&0&0
\end{array}
\right)
\ee
for $k+2~ mode~4 =3$.

One can prove that the matrices in (\ref{newa}) also give rise to
$d$-computable states:
$$
|A_{2^{k+1}}A_{2^{k+1}}^\dag|
=[(c^2+ad-\sum_{i=1}^{k}b_ic_i)(c^2+ad-\sum_{i=1}^{k}b_ic_i)^*]^{2^k},
$$
$$
\ba{rcl}
|A_{2^{k+1}}A_{2^{k+1}}^\dag-\lambda Id_{2^{k+1}}|
&=&\displaystyle[\lambda^2-(aa^\ast+2cc^\ast+dd^\ast
+\sum_{i=1}^{k}b_ib_i^\ast+\sum_{i=1}^{k}c_ic_i^\ast)\lambda\\[4mm]
&&\displaystyle+(c^2+ad
-\sum_{i=1}^{k}b_ic_i)(c^2+ad-\sum_{i=1}^{k}b_ic_i)^*]^{2^k}.
\ea
$$
The entanglement of formation for a density
matrix with decompositions in these states is also given by a formula
of the form (\ref{d2k1}).

In addition, the results obtained above may be used to solve
linear equation systems, e.g., in the analysis of data bank,
described by $A{\bf x}={\bf y}$, where $A$
is a $2^{k}\times 2^{k}$ matrix, $k\in\Nb$, ${\bf x}$ and ${\bf
y}$ are $2^{k}$-dimensional column vectors.
When the dimension $2^{k}$ is large, the standard
methods such as Gauss elimination to solve $A{\bf x}={\bf y}$
could be not efficient. From our Lemma 3, if the
matrix $A$ is of one of the following forms: $A_{2^k}$,
$B_{2^k}A_{2^k}$, $A_{2^k}^t$ or $A_{2^k}^tB_{2^k}^t$, the
solution ${\bf x}$ can be obtained easily by applying the matrix
multiplicators. For example, $A_{2^k}{\bf x}={\bf y}$ is solved by
$$
{\bf x}=\frac{1}{[A_{2^k}]}(A_{2^k}J_{2^k})^t J_{2^k}{\bf y}.
$$
The solution to $B_{2^k}A_{2^k}{\bf x}={\bf y}$ is given by
$$
{\bf x}=\frac{1}{b_k[A_{2^k}]}(A_{2^k}J_{2^k})^t J_{2^k}{\bf y}.
$$

We have presented a kind of construction for a class of special matrices
with at most two different eigenvalues. This class of matrices
defines a special kind of $d$-computable states.
The entanglement of formation for these $d$-computable states
is a monotonically increasing function
of a the generalized concurrence. From this generalized concurrence
the entanglement of formation for a large class of density matrices
whose decompositions lie in these $d$-computable quantum states is obtained
analytically. Besides the relations to the quantum entanglement,
the construction of $d$-computable states has its own mathematical interests.

\end{document}